\documentclass[aps,prd,preprint]{revtex4}

\usepackage{graphicx}
\usepackage{psfrag}
\usepackage{epsfig}

\begin{document}

\title{Pion Form Factor in the $k_T$ Factorization Formalism}
\author{Tao Huang$^{1,2}$\footnote{email:
huangtao@mail.ihep.ac.cn}, Xing-Gang Wu$^{2}$\footnote{email:
wuxg@mail.ihep.ac.cn} and Xing-Hua
Wu$^2$\footnote{email:xhwu@mail.ihep.ac.cn}}
\address{$^1$CCAST(World Laboratory), P.O.Box 8730, Beijing 100080,
P.R.China,\\
$^2$Institute of High Energy Physics, Chinese Academy of Sciences,
P.O.Box 918(4), Beijing 100039, China.\footnote{Mailing address}}

\begin{abstract}
Based on the light-cone (LC) framework and the $k_T$ factorization
formalism, the transverse momentum effects and the different
helicity components' contributions to the pion form factor
$F_{\pi}(Q^2)$ are recalculated. In particular, the contribution
to the pion form factor from the higher helicity components
($\lambda_1+\lambda_2=\pm 1$), which come from the spin-space
Wigner rotation, are analyzed in the soft and hard energy regions
respectively. Our results show that the right power behavior of
the hard contribution from the higher helicity components can only
be obtained by fully keeping the $k_T$ dependence in the hard
amplitude, and that the $k_T$ dependence in LC wave function
affects the hard and soft contributions substantially. As an
example, we employ a model LC wave function to calculate the pion
form factor and then compare the numerical predictions with the
experimental data. It is shown that the soft contribution is
less important at the intermediate energy region.\\


\end{abstract}
\maketitle

\section{Introduction}

In the perturbative QCD (PQCD) theory, the hadronic distribution
amplitudes and structure functions which enter exclusive and
inclusive processes via the factorization theorems at high
momentum transfer can be determined by the hadronic wave
functions, and therefore they are the underlying links between
hadronic phenomena in QCD at large distances (nonperturbative) and
small distance (perturbative). However we require a conceptual
framework within which the connection between the hadrons and
their constituents can be made precise. A particularly convenient
and intuitive framework is based upon the Fock state decomposition
of hadronic states which arises naturally in the `light-cone
quantization'~\cite{lbhm,lb}. A light-cone (LC) wave function is a
localized (i.e. normalizable) stationary solution of the LC
schr\"{o}dinger equation $i\partial |\Psi(\tau)\rangle
=H_{LC}|\Psi(\tau)\rangle$ that describes the evolution of a state
$|\Psi(\tau)\rangle$ on the LC time $\tau\equiv x^{+}=x^{0}+x^{3}$
in physical light cone gauge $A^+=A^0+A^3=0$, which is conjugate
to the LC Hamiltonian $H_{LC}\equiv P^{-}=P^{0}-P^{3}$. The LC
wave functions are the amplitudes $\Psi_n(x_i,{\bf k}_{\perp
i},\lambda_i)$ to find $n$ particles (quarks, antiquarks and
gluons) with momenta $k_i$ in a pion of momentum $P$, i.e.,
$\Psi(k_1,\cdots,k_n;P)\equiv \langle n|\pi\rangle=\langle
k_1,\cdots,k_n|\pi(P)\rangle$, where $x_i=k_i^+/P^+$, with $\sum_i
x_i=1$, is the LC momentum fraction of the $i$-th quark or gluon
in the $n-$particle Fock state.

An important issue, which has to be addressed when applying PQCD
to exclusive processes, is how to implement factorization, i.e.,
separate perturbative contributions from those intrinsic to the
bound-state wave function. Both collinear and $k_T$ factorization
are the fundamental tools of PQCD. If there is no end-point
singularity developed in a hard amplitude, collinear factorization
works. If such singularity occurs, indicating the breakdown of
collinear factorization, (one may find a concrete example in the
semi-leptonic decay $B\rightarrow \pi l\bar{\nu}$\cite{shb}), then
$k_T$ factorization should be employed. In the $k_T$ factorization
formula, by retaining the dependence on the parton transverse
momentum $k_T$ and resuming the resultant double logarithms
$\alpha_s\ln^2k_T$ into a Sudakov form factor, such singularity
does not exist\cite{liyu}. Since the $k_T$ factorization theorem
has been proposed\cite{lbhm,lb,lis}, it has been widely applied to
various processes. Until recently, a better proof of the $k_T$
factorization theorem for exclusive processes in PQCD has been
provided by M. Nagashima and H.N. Li\cite{nli}. Their starting
point is that the on-shell valence partons carry longitudinal
momenta initially, and then acquire $k_T$ through collinear gluon
exchanges before participating in hard scattering. A hard
amplitude, derived from the parton level amplitudes with the
gauge-invariant and infrared divergent meson wave function being
subtracted, is then gauge-invariant and infrared-finite. Through
this way, they demonstrated that all the physical quantities from
the $k_T$ factorization theorem are gauge-invariant. Therefore for
the pion form factor, when in the energy region that PQCD is
applicable, we can take the following factorization
formula\cite{lbhm,lis,pi2pi,jps,cchm},
\begin{equation}
\label{fpieff} F_{\pi}(Q^2)=\sum_{n,m,\lambda_{i,j}}\int[dx_id{\bf
k}_{i\perp}]_{n}[dy_jd{\bf l}_{j\perp}]_{m} \psi^*_{n}(x_i,{\bf
k}_{i\perp},\lambda_i;\mu)T_{n m}(x_i,{\bf k}_{i\perp};y_j,{\bf
l}_{j\perp};{\bf q}_{i\perp};\mu)\psi_{m}(y_j,{\bf
l}_{j\perp},\lambda_j;\mu)
\end{equation}
where summation over all helicities ($\lambda_{i}$,$\lambda_j$)
and $n,\ m$ extends over the low momentum states only, and
$T_{n,m}$ are the partonic matrix elements of the effective
current operator. $\mu$ is the energy scale separating the
perturbative from the non-perturbative region, and in order for
the perturbative approach to make sense, $\mu$ has to be much
larger than $\Lambda_{QCD}$ so that $\alpha_s(\mu)$ is small.

We notice that although most of the calculations show that PQCD is
self-consistent and applicable to the exclusive processes at
currently experimental accessible energy region, the numerical
predictions for the pion form factor are much smaller than the
experimental data. There are two possible explanations: one is
that the non-perturbative contribution will be important in this
region; the other is that the non-leading order contribution in
perturbative expansions may be also important in this region. To
make choice between those two possible explanations one needs to
analyze the non-leading contributions which come from higher-twist
effect\cite{twist}, higher order in $\alpha_s$\cite{field}, and
higher Fock states\cite{pi2pi} {\it etc.}. Employing the modified
factorization expression for the pion form factor proposed by Li
and Sterman \cite{lis}, Refs.\cite{pt} considered the effect of
the transverse momentum ($k_T$) in the wave function and found
that the transverse momentum in the wave function plays the role
to suppress perturbative prediction. V.M. Braun {\it
etal.}\cite{bkm} gave a detailed quantitative analysis of the pion
form factor in the region of intermediate momentum transfers in
the LC sum rule approach and they observed a strong numerical
cancellation between the soft contribution and the power
suppressed hard contribution of higher twist and then the total
non-perturbative correction to the usual PQCD result to be of
order $30\%$ for $Q^2\sim 1GeV^2$.

One of the other sources which may provide non-leading
perturbative contribution is the higher helicity components in the
LC wave function \cite{mh,wk,cchm}. However, the results for the
contribution coming from higher helicity components to the pion
form factor in the high energy region are conflicting in
literature\cite{mh,wk}. The hard scattering amplitude for the
higher helicity components of the pion form factor at the leading
order of $\alpha_s(Q^2)$ was given by Ref.\cite{cchm} in the LC
framework. In present paper, we recalculate all the helicity
components' contributions to the pion form factor within the LC
PQCD framework, which is consistent with the using of LC wave
function. Our calculation keeps the transverse momentum dependence
fully in the hard scattering amplitude, i.e. such dependence is
kept in both the quark propagator and the gluon propagator, and
the resultant expression gives the right power behavior of the
hard contribution from the higher helicity components as $Q^2$
goes to large energy region. Furthermore, we carry out the
numerical calculations for the hard and the soft parts of all the
helicity components' contributions. In order to explain our
picture and to clarify the difference between Ref.\cite{mh} and
Ref.\cite{wk}, we employ a model LC wave function with reasonable
constraints. We show that it is substantial to take $k_T$
dependence in the wave function into account and to keep the
transverse momentum dependence fully in the hard scattering
amplitude in the $k_T$ factorization formalism within the LC
framework.

The purpose of this paper is to reanalyze the effects coming from
the higher-helicity components of the pion wave function within
the framework of LC PQCD and the $k_T$ factorization formalism,
then give a comparative study on the contributions from different
helicity components within the soft and the hard region
respectively. In section II, based on the $k_T$ factorization
formula, the hard scattering amplitude is given within the LC
framework. In section III, with a model LC wave function, the hard
contributions from different helicity components of pion are
analyzed. Section IV is devoted to give a discussion of the soft
part contribution, especially on the contribution from different
helicity components. Conclusion and a brief summary are presented
in the final section.

\section{Hard scattering amplitude with $k_T$ dependence}

In the light cone quantization, the pion form factor can generally
be expressed by using the Drell-Yan-West ($q^+=0$) frame\cite{dy},
\begin{equation}\label{basic:form}
F_{\pi} (Q^2) = \hat\Psi \otimes \hat\Psi =\sum_{n,\lambda_i}\int
[dx_i] [dk_{i\perp}]_n \Psi^*_n(x_i,k_{i\perp},\lambda_i)
\Psi_n(x_i,k_{i\perp}+\delta_iq_\perp,\lambda_i), \label{fpi0}
\end{equation}
where the summation extends over all quark/gluon Fock states which
have a non-vanishing overlap with the pion, $\Psi_n$ are the
corresponding wave functions which describe both the low and the
high momentum partons, $[dx_i][dk_{i\perp}]_n$ is the relativistic
measure within the $n$-particle sector and $\delta_i=(1-x_i)$ or
$(-x_i)$ depending on whether $i$ refers to the struck quark or a
spectator, respectively. From Eq.(\ref{basic:form}) and the $k_T$
factorization formula Eq.(\ref{fpieff}), at higher momentum
transfer, the hard contribution to the pion form factor can be
written as\cite{mh,lis,jps,cchm}
\begin{equation}
\label{factor} F_{\pi}(Q^2)=\int [dx][dy][d^2{\bf
k_{\perp}}][d^2{\bf l_{\perp}}] \psi^{*(1-x)Q}(x,{\bf
k_{\perp}},\lambda) T_{H}(x,y,{\bf q_{\perp},{\bf k_{\perp}},{\bf
l_{\perp}},\lambda,\lambda^{\prime}})\psi^{(1-y)Q}(y,{\bf
l_{\perp}},\lambda^{\prime})+ \cdots,
\end{equation}
where the ellipses represent the higher Fock states'
contributions, $[dx]=dx_1d_2\delta(1-x_1-x_2)$ and $[d^2{\bf
k_{\perp}}]=d^2{\bf k_{\perp}}/16\pi^3$. $\psi^{((1-x)Q)}(x,{\bf
k}_\perp,\lambda)$ is the valence Fock state LC wave function with
helicity $\lambda$ and with a cut-off on $|{\bf k}_\perp|$ that is
of order $(1-x)Q$. Such a cut-off on $|{\bf k}_\perp|$ is
necessary to insure that the wave function is only responsible for
the lower momentum region. $T_H$ contains all two-particle
irreducible amplitudes for $\gamma^*+q\bar{q}\to q\bar{q}$ and
should be calculated from the time-ordered diagrams in LC PQCD. In
the light cone gauge $(A^+=0)$, the nominal power law contribution
to $F_{\pi}(Q^2)$ as $Q\rightarrow \infty$ is $F_{\pi}(Q^2)\sim
1/(Q^2)^{n-1}$\cite{brodsky}, under the condition that $n$ quark
or gluon constituents are forced to change direction. Thus only
the $q\bar{q}$ component of $\psi^{((1-x)Q)}(x,{\bf
k}_\perp,\lambda)$ contributes at the leading $1/Q^2$.

\begin{figure}
\setlength{\unitlength}{1mm}
\begin{picture}(80,80)(30,30)
\put(-5,-40){\includegraphics{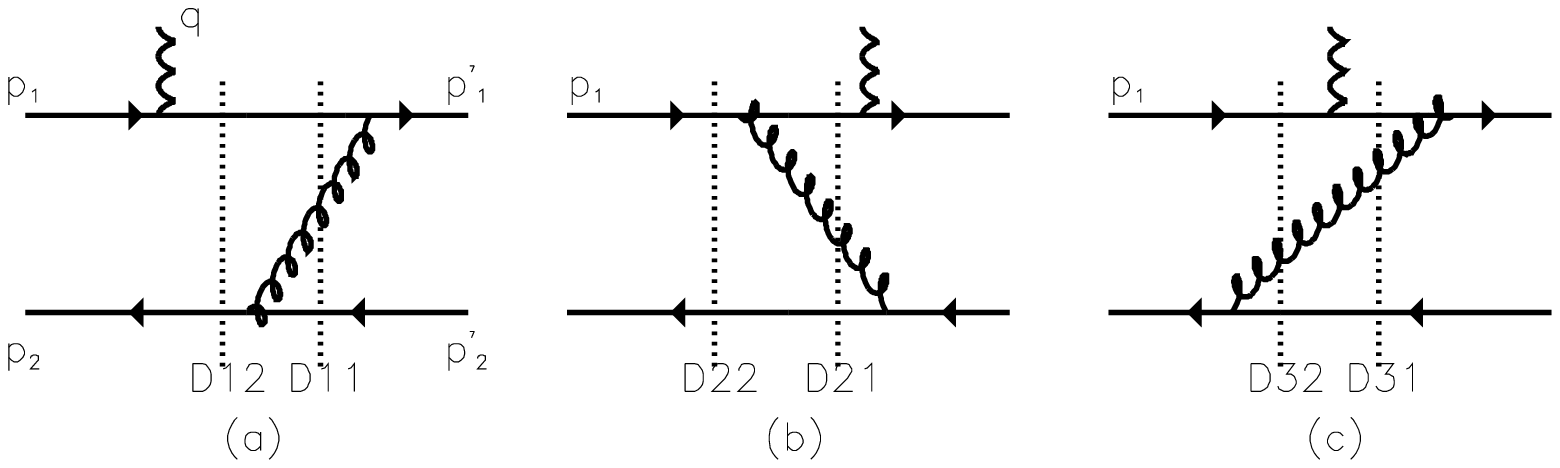}}
\put(-5,-80){\includegraphics{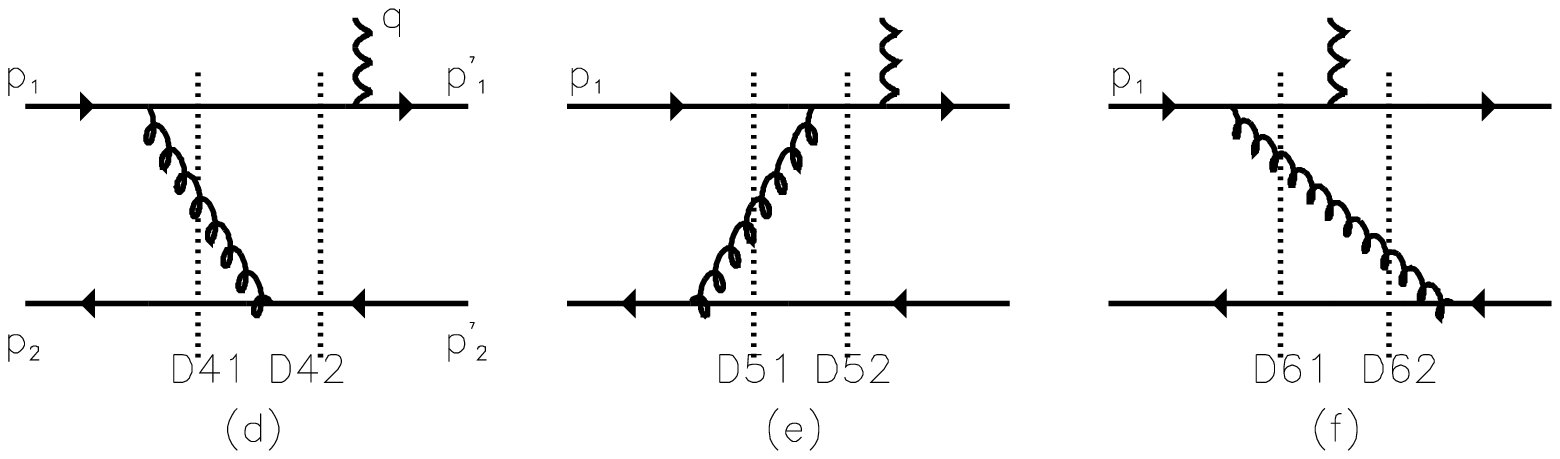}}
\end{picture}
\caption{Six leading order time-ordered Feynman diagrams for the
hard scattering amplitude, where $p_1=(x_1,\mathbf{k}_{\perp})$,
$p_2=(x_2,-\mathbf{k}_{\perp})$,
$p_1^{\prime}=(y_1,y_1\mathbf{q}_{\perp}+\mathbf{l}_{\perp})$,
$p_2^{\prime}=(y_2,y_2\mathbf{q}_{\perp}-\mathbf{l}_{\perp})$. }
\label{figpi1}
\end{figure}

The lowest-order contribution for the hard scattering amplitude
$T_H$ comes from the one-gluon exchange shown in Fig.\ref{figpi1}.
To simplicity our notations, we separate the spin-space wave
function $\chi^K(x,{\bf k}_{\perp},\lambda)$ out from the whole LC
wave function, i.e., $\psi^{(1-x)Q}(x,{\bf
k_{\perp}},\lambda)\to\chi^K(x,{\bf
k}_{\perp},\lambda)\varphi^{(1-x)Q}(x,{\bf k_{\perp}},\lambda)$
and then combined the spin-space wave function $\chi^K(x,{\bf
k}_{\perp},\lambda)$ into the original $T_H$ to form a new one,
i.e.,
\begin{eqnarray}
T_H &=& \xi_1 T_H^{(\lambda_1+\lambda_2=0)}(\uparrow\downarrow
\rightarrow\uparrow\downarrow)+ \xi_1
T_H^{(\lambda_1+\lambda_2=0)}(\downarrow\uparrow
\rightarrow\downarrow\uparrow)+ \xi_2 T_H^{(\lambda_1+\lambda_2=
1)}(\uparrow\uparrow\rightarrow\uparrow\uparrow)+\nonumber\\
& & \xi_2^* T_H^{(\lambda_1+\lambda_2=- 1)}(\downarrow\downarrow
\rightarrow\downarrow\downarrow)\ ,
\end{eqnarray}
where $\lambda_{1,2}$ are the helicities for the (initial or
final) pion's two constitute quarks respectively, $\xi_1=
\frac{m^2}{2[m^2+\mathbf{k}^2_{\perp}]^{1/2}
[m^2+\mathbf{l}^2_{\perp}]^{1/2}}$ and
$\xi_2=\frac{\mathbf{k}_{\perp}\cdot
\mathbf{l}_{\perp}+i(\mathbf{k}_{\perp}\times
\mathbf{l}_{\perp})}{2[m^2+\mathbf{k}^2_{\perp}]^{1/2}
[m^2+\mathbf{l}^2_{\perp}]^{1/2}}$ are two coefficients derived
from $\chi^K(x,{\bf k}_{\perp},\lambda)$. The spin space wave
function $\chi^K(x,{\bf k}_{\perp},\lambda)$ which comes from the
spin space Wigner rotation can be found in Ref.\cite{hms}. Because
both photon and gluon are vector particles, the quark helicity is
conserved at each vertex in the limit of vanishing quark
mass\cite{lb}. Hence there is no hard-scattering amplitude with
quark and antiquark helicities being changed.

To simplify the hard scattering amplitude, we adopt the standard
momentum assignment at the ``infinite-momentum" frame~\cite{lb},
\begin{equation}
P_{\pi}=(P^{+},P^{-},{\bf P_{\perp}})=(1,0,{\bf 0}_{\perp}),\;\;\;
q=(0,q^{2}_{\perp},{\bf q}_{\perp}),
\end{equation}
where $P^+$ is arbitrary because of Lorentz invariance and the
square of the momentum transfer
$Q^2=-q^{2}=\mathbf{q}^{2}_{\perp}$. Using $D$ to denote the
``energy-denominator" in the 6 Feynman diagrams ($x^+$-ordered
diagrams), all the needed ``energy-denominators" are listed in the
following\cite{cchm},
\begin{eqnarray}
D_{11} &=&-\frac{(x_2{\bf q}_\perp+{\bf k}_\perp)^2}{x_1x_2}
       -\frac{[y_2(x_2{\bf q}_\perp+{\bf k}_\perp)-x_2{\bf l}_\perp]^2}
         {x_2y_2(y_1-x_1)}\ ,\\
D_{12} &=& -\frac{(x_2{\bf q}_\perp+{\bf k}_\perp)^2}{x_1x_2}\ ,\\
D_{21} &=& -\frac{{\bf l}_\perp^2}{y_1y_2}
       +\frac{[y_2(x_2{\bf q}_\perp+{\bf k}_\perp)-x_2{\bf l}_\perp]^2}
         {x_2y_2(y_1-x_1)}\ , ~~~ D_{22}=D_{12}\ ,\\
D_{32} &=& -\frac{{\bf k}_\perp^2}{x_1x_2}
       -\frac{[y_2(x_2{\bf q}_\perp+{\bf k}_\perp)-x_2{\bf l}_\perp]^2}
         {x_2y_2(y_1-x_1)}\ ,~~~ D_{31}=D_{11}\ ,\\
D_{i+3,j}&=& D_{i,j}(x\leftrightarrow y,\;
\mathbf{k}_{\perp}\leftrightarrow -\mathbf{l}_{\perp}),
\;\;\;(i=1,2,3;\;j=1,2) \ ,
\end{eqnarray}
where the last equation comes from the charge symmetry. With the
help of the above equations, the hard scattering amplitude can be
shortly expressed as,
\begin{equation}
T^{(\lambda_1+\lambda_2)}_{H}=g^2C_{F}\bigg(T^{(\lambda_1+\lambda_2)}_{a}
+T^{(\lambda_1+\lambda_2)}_{b}+T^{(\lambda_1+\lambda_2)}_{c}\bigg)+\left\{
\begin{array}{lcc}
x&\leftrightarrow&y\\
{\bf k}_\perp&\leftrightarrow&-{\bf l}_\perp
\end{array}\right\},
\end{equation}
where the three terms in the parentheses, which correspond to
Fig.\ref{figpi1}.a, Fig.\ref{figpi1}.b and Fig.\ref{figpi1}.c
respectively, can be written as
\begin{eqnarray}
T^{(\lambda_1+\lambda_2)}_{a}&=&\frac{N^{(\lambda_1+\lambda_2)}}
{D_{11}D_{12}}\frac{\theta(y_{1} -x_{1})}{y_1-x_1}+T^{in}_{a},~~~
T^{in}_{a}= \frac{-4}{D_{12}}\frac{\theta(y_{1}-x_{1})}{(y_1-x_1)^2},\\
T^{(\lambda_1+\lambda_2)}_{b}&=&\frac{N^{(\lambda_1+\lambda_2)}}
{D_{21}D_{22}}\frac{\theta(x_{1} -y_{1})}{x_1-y_1}+T^{in}_{b},~~~
T^{in}_{b}= \frac{-4}{D_{22}}\frac{\theta(x_{1}-y_{1})}{(y_1-x_1)^2},\\
T^{(\lambda_1+\lambda_2)}_{c}&=&\frac{N^{(\lambda_1+\lambda_2)}}
{D_{31}D_{32}}\frac{\theta(y_{1} -x_{1})}{y_1-x_1}.
\end{eqnarray}
Here $T^{in}_{a}$ and $T^{in}_{b}$ represent the contributions
from the instantaneous diagrams in the light-cone PQCD. The
numerator $N^{(\lambda_1+\lambda_2)}$ is the sum of some spinors
and $\gamma$ matrixes and for the usual helicity components
$(\lambda_1+\lambda_2=0)$,
\begin{eqnarray}
N^{(\lambda_1+\lambda_2=0)} &=&
-\mathbf{q}^2_{\perp}\left(\frac{x_2(x_1x_2+y_1y_2)}{x_1(y_1-x_1)^2}\right)
-\mathbf{k}_{\perp}^2\left(\frac{(x_1x_2+y_1y_2)}{x_1x_2(y_1-x_1)^2}\right)
-\mathbf{l}_{\perp}^2\left(\frac{(x_1x_2+y_1y_2)}{y_1y_2(y_1-x_1)^2}\right)\nonumber\\
& & \!\!\!\!\!\!\!-(2{\bf q}_{\perp}\cdot {\bf
k}_{\perp})\left(\frac{(x_1x_2+y_1y_2)}{x_1(y_1-x_1)^2}\right)+({\bf
q}_{\perp}\cdot {\bf l}_{\perp})\left(\frac{(x_1x_2+
y_1y_2)(x_2y_1+x_1y_2)}{x_1y_1y_2(y_1-x_1)^2}\right)+\nonumber\\
& & \!\!\!\!\!\!\! ({\bf k}_{\perp}\cdot {\bf
l}_{\perp})\left(\frac{(x_1x_2+
y_1y_2)(x_2y_1+x_1y_2)}{x_1x_2y_1y_2(y_1-x_1)^2}\right)\pm i
\left( \frac{(x_2-y_1)(\mathbf{l}_{\perp} \times
(\mathbf{k}_{\perp}+x_2 \mathbf{q}_{\perp})} {x_1 x_2 y_1
y_2}\right)
\end{eqnarray}
where the plus sign corresponds to
$(\uparrow\downarrow\rightarrow\uparrow\downarrow)$ and the minus
sign corresponds to $(\downarrow\uparrow \rightarrow\downarrow
\uparrow)$. And for the higher helicity components
$(\lambda_1+\lambda_2=\pm 1)$,
\begin{eqnarray}
N^{(\lambda_1+\lambda_2=\pm 1)} &=&
-\mathbf{q}^2_{\perp}\left(\frac{x_2(x_2y_1+x_1y_2)}{x_1(y_1-x_1)^2}\right)
-\mathbf{k}_{\perp}^2\left(\frac{(x_2y_1+x_1y_2)}{x_1x_2(y_1-x_1)^2}\right)
-\mathbf{l}_{\perp}^2\left(\frac{(x_2y_1+x_1y_2)}{y_1y_2(y_1-x_1)^2}\right)\nonumber\\
& & -(2{\bf q}_{\perp}\cdot {\bf k}_{\perp})\left(\frac{(x_2y_1+
x_1y_2)}{x_1(y_1-x_1)^2}\right)+({\bf q}_{\perp}\cdot {\bf
l}_{\perp})\left(\frac{(x_2y_1+
x_1y_2)^2}{x_1y_1y_2(y_1-x_1)^2}\right)+\nonumber\\
& &({\bf k}_{\perp}\cdot {\bf
l}_{\perp})\left(\frac{(x_2y_1+x_1y_2)^2}{x_1x_2y_1y_2
(y_1-x_1)^2}\right)\pm i \left( \frac{(\mathbf{l}_{\perp} \times
(\mathbf{k}_{\perp}+x_2 \mathbf{q}_{\perp})} {x_1 x_2 y_1
y_2}\right)\ ,
\end{eqnarray}
where the plus sign corresponds to $\lambda_1+\lambda_2=1$
($\uparrow\uparrow\rightarrow\uparrow\uparrow$) and the minus sign
corresponds to $\lambda_1+\lambda_2=-1$
($\downarrow\downarrow\rightarrow\downarrow\downarrow$).

In order to further simplify the hard scattering amplitude, we
adopt the following two prescriptions: 1) It is pointed out in
Ref.~\cite{jps} that when one concerns with the effect from the
intrinsic transverse momenta, the terms proportional to the
``bound energies" of the pions in the initial and final states
{\it i.e.} $\sim {\bf k}_\perp^2/(x_1x_2)$ and $\sim {\bf
l}_\perp^2/(y_1y_2)$ can be ignored to avoid the involvement of
the higher Fock states' contributions\footnote{As the transverse
momenta ${\bf k}_\perp$ and ${\bf l}_\perp$ are included, it is
necessary to take into account the contributions from higher Fock
states to satisfy the gauge-invariance, since the covariant
derivative $D_\mu=\partial_\mu+igA_\mu$ makes both transverse
momenta ${\bf k}_\perp$, ${\bf l}_\perp$ and the transverse gauge
degree $g{\bf A}_\perp$ be of the same order \cite{jps}.}. 2)
Notice that in the factorization expression for the pion form
factor Eq.(\ref{factor}), we have ${\bf k}_\perp^2\ll{\bf
q}_\perp^2$ and ${\bf l}_\perp^2\ll{\bf q}_\perp^2$. Hence when
calculating $T_H$ to the next-to-leading order in $1/Q$, we can
safely neglect the terms such as ${\bf k}_\perp^2/{\bf q}_\perp^2$
and ${\bf l}_\perp^2/{\bf q}_\perp^2$ in both the ``energy
denominators" and the numerator $N^{(\lambda_1+\lambda_2)}$.

The natural variable to make a separation of perturbative
contributions from those intrinsic to the bound-state wave
function is the LC energy in the LC perturbative
expansion\cite{bhl,jps}. Under such condition, the two energy flow
$\left(-\frac{(x_2 \mathbf{q}_{\perp}+ \mathbf{k}_{\perp})^2}{x_1
x_2}\right)$ and $\left(-\frac{(y_2(x_2
\mathbf{q}_{\perp}+\mathbf{k}_{\perp})-x_2 l_{\perp})^2}{x_2 y_2
(y_1-x_1)}\right)$ in the gluon propagator should be large,
otherwise we can't apply the PQCD, i.e., $$ (x_2
\mathbf{q}_{\perp}+ \mathbf{k}_{\perp})^2 \gg \langle
\mathbf{k}^2_{\perp}\rangle\sim \bar{\Lambda}^2 $$ and
$$(y_2(x_2 \mathbf{q}_{\perp}+\mathbf{k}_{\perp})-x_2 \mathbf{l}_{\perp})^2\gg \langle
\mathbf{k}^2_{\perp}\rangle,\;\; \langle
\mathbf{l}^2_{\perp}\rangle \sim \bar{\Lambda}^2 $$ are the
conditions which make the PQCD applicable, where $\bar{\Lambda}$,
being of ${\cal O}(\Lambda_{QCD})$, represents a hadronic scale.

Applying the above two prescriptions, we finally obtain
\begin{equation}
T_H=T_{H}^{(\lambda_1+\lambda_2=0)}+
T_{H}^{(\lambda_1+\lambda_2=\pm 1)}
\end{equation}
with
\begin{eqnarray}\label{hardamp1}
T_{H}^{(\lambda_1+\lambda_2=0)}&=&\frac{16\xi_1  \pi {C_F}
{\alpha_s(Q^2)}}{ ( 1- x )( 1-y )x y} \times((( x-1
)\mathbf{q_\perp}^2 -2\mathbf{k_\perp}\cdot \mathbf{q_\perp})( 2
\mathbf{l_\perp}\cdot\mathbf{q_\perp} +
(y -1)\mathbf{q_\perp}^2 ))^{-1}\nonumber\\
& & ((x-1) ( 2 \mathbf{l_\perp} \cdot\mathbf{q_\perp} +   (y
-1)\mathbf{q_\perp}^2 )-2 (y -1)
\mathbf{k_\perp} \cdot\mathbf{q_\perp})^{-1}\times\nonumber\\
& & \Bigg( 2 ( y-1) y ( 1 - y + x ( 2 y-1 ) )
{(\mathbf{k_\perp}\cdot\mathbf{q_\perp})}^2 + ( x-1) x (
2\mathbf{l_\perp}\cdot\mathbf{q_\perp} + ( y-1
)\mathbf{q_\perp}^2)\cdot\nonumber\\
& & ( ( 1 - y + x ( 2y-1 ) )
(\mathbf{l_\perp}\cdot\mathbf{q_\perp})+ 2 ( x-1 ) ( y-1) y
\mathbf{q_\perp}^2 ) - \nonumber\\
& & ( x-1) ( y-1) y (\mathbf{k_\perp}\cdot\mathbf{q_\perp})\cdot(
8 x (\mathbf{l_\perp}\cdot\mathbf{q_\perp}) + ( 1 - y + x ( 6y-5))
\mathbf{q_\perp}^2 ) \Bigg),
\end{eqnarray}
and
\begin{eqnarray}\label{hardamp2}
T_{H}^{(\lambda_1+\lambda_2=\pm 1)}&=& \frac{8 (\xi_2+\xi_2^*)\pi
A^2 {C_F} {\alpha_s(Q^2)}}{ ( 1- x )( 1-y )x y } \times((( x-1
)\mathbf{q_\perp}^2-2\mathbf{k_\perp}\cdot\mathbf{q_\perp}) (2
\mathbf{l_\perp}\cdot\mathbf{q_\perp} +  (y -1)
\mathbf{q_\perp}^2 ))^{-1}\nonumber\\
& & ((x-1) ( 2 \mathbf{l_\perp} \cdot\mathbf{q_\perp} +   (y
-1)\mathbf{q_\perp}^2 )-2 (y -1) \mathbf{k_\perp}
\cdot\mathbf{q_\perp})^{-1}\Bigg( 2( x-1) x {(\mathbf{l_\perp}
\cdot\mathbf{q_\perp})}^2+ \nonumber\\
& & ( y-1) ( 2y{(\mathbf{k_\perp} \cdot\mathbf{q_\perp})}^2 + (
x-1) (x(\mathbf{l_\perp} \cdot\mathbf{q_\perp})- y
(\mathbf{k_\perp}\cdot\mathbf{q_\perp})) \mathbf{q_\perp}^2)
\Bigg).
\end{eqnarray}
After doing a simple transformation, one may find that the
obtained hard scattering amplitude for the higher helicity
components $(\lambda_1+\lambda_2=\pm 1)$ coincides well with the
one obtained in Ref.\cite{cchm}, while the hard scattering
amplitude for the usual helicity components
$(\lambda_1+\lambda_2=0)$ is different from others after including
all the $k_T$ dependence in the LC PQCD framework. Due to the
complicated integral in Eq.(\ref{factor}), Ref.\cite{cchm} didn't
give the numerical results for the higher helicity contribution of
the pion form factor. We will apply the VEGAS program\cite{vegas}
to evaluate the hard contribution in the next sections.

From Eqs.(\ref{hardamp1},\ref{hardamp2}), ignoring the $k_{\perp}$
dependence, we obtain
\begin{eqnarray}\label{approxh}
& & T^{(\lambda_1+\lambda_2=0)}_{H}=
T^{(\lambda_1+\lambda_2=0)}_{H}(\uparrow\downarrow\rightarrow\uparrow\downarrow)+
T^{(\lambda_1+\lambda_2=0)}_{H}(\downarrow\uparrow
\rightarrow\downarrow \uparrow)=\frac{16\pi C_F
\alpha_{s}(Q^2)}{x_2y_2 Q^2},\nonumber\\
& & T^{(\lambda_1+\lambda_2=\pm 1)}_{H}=
T^{(\lambda_1+\lambda_2=1)}_{H}(\uparrow\uparrow\rightarrow\uparrow\uparrow)+
T^{(\lambda_1+\lambda_2=-1)}_{H}(\downarrow\downarrow
\rightarrow\downarrow \downarrow)=0\ .
\end{eqnarray}
It can be found from Eqs.(\ref{hardamp2},\ref{approxh}) that the
leading contribution from the higher helicity components is of
order $1/Q^4$, which is next-to-leading contribution compared to
that of the ordinary helicity components.

\section{Hard contribution to the pion form factor}

In order to get the hard contribution for the pion form factor
from Eq.(\ref{factor}), we need to know the soft hadronic wave
function. Several important non-perturbative approaches have been
developed to provide the theoretical predictions for the hadronic
wave functions\cite{sumrule,lattice,hms,kr,bkm,bf,bhl}. One useful
way is to use the approximate bound state solution of a hadron in
terms of the quark model as the starting point for modelling the
hadronic valence wave function. The Brodsky-Huang-Lepage (BHL)
prescription~\cite{bhl} of the hadronic wave function is in fact
obtained in this way by connecting the equal-time wave function in
the rest frame and the wave function in the infinite momentum
frame. In Ref.~\cite{hms}, based on the BHL prescription, a
revised LC quark model wave function has been raised that can give
both the approximate asymptotic distribution amplitude and the
reasonable valence state structure function which does not exceed
the pion structure function data simultaneously. So in the present
paper, we will use this revised LC quark model wave function for
our latter discussions, i.e.
\begin{equation}\label{wave}
 \Psi(x,{\bf k}_{\perp})=\varphi_{\mathrm{BHL}}(x,{\bf k}_{\perp})
 \chi^K(x,{\bf k}_{\perp})=A \exp[-\frac{{\bf k}_{\perp}^2
  +m^2}{8{\beta}^2x(1-x)}]\chi^K(x,{\bf k}_{\perp}),
\end{equation}
with the parameters, the normalization constant $A$, the harmonic
scale $\beta$ and the quark mass $m$ to be determined. With the
help of the model wave function, from Eq.(\ref{factor}), we can
obtain the leading-twist hard part contribution to the pion form
factor. From Eq.(\ref{hardamp1}) we obtain the contribution from
the usual helicity components $(\lambda_1+\lambda_2=0)$,
\begin{eqnarray}\label{hel0}
F^{(\lambda_1+\lambda_2=0)}_{\pi}(Q^2)&=&\int dx dy[d^2{\bf
k_{\perp}}][d^2{\bf l_{\perp}}] \frac{8 \pi A^2 \xi_1 {C_F}
{\alpha_s(Q^2)}}{ ( 1- x )( 1-y )x
y}\times\exp\left(-\frac{\frac{m^2 +\mathbf{k_{\perp}}^2} {x(1-x)}
+ \frac{m^2 + \mathbf{l_{\perp}}^2}{y(1-y)}}{8 {\beta
}^2}\right)\times\nonumber\\
& & \left( \frac{x(x+y-2xy-1)} {( 1-x) \mathbf{q_{\perp}}^2+2
\mathbf{q_{\perp}} \cdot \mathbf{k_{\perp}}} +\frac{y(x+y-2xy-1)}
{(1-y)\mathbf{q_{\perp}}^2-2\mathbf{q_{\perp}}\cdot
\mathbf{l_{\perp}}}+\right.\nonumber\\
& & \left. \frac{x+y-x^2-y^2}{2 (1-y) \mathbf{q_{\perp}}\cdot
\mathbf{k_{\perp}}-2(1-x)\mathbf{q_{\perp}}\cdot
\mathbf{l_{\perp}}+ (1-x)(1-y)\mathbf{q_{\perp}}^2} \right) ,
\end{eqnarray}
and from Eq.(\ref{hardamp2}) we obtain the contribution from
higher helicity components $(\lambda_1+\lambda_2=\pm 1)$,
\begin{eqnarray}\label{hel1}
F^{(\lambda_1+\lambda_2=\pm 1)}_{\pi}(Q^2)&=&\int dx dy[d^2{\bf
k_{\perp}}][d^2{\bf l_{\perp}}] \frac{4 \pi A^2 (\xi_2+\xi_2^*)
{C_F} {\alpha_s(Q^2)}}{ ( 1- x )( 1-y )x y
}\times\exp\left(-\frac{\frac{m^2 +\mathbf{k_{\perp}}^2} {x(1-x)}
+ \frac{m^2 + \mathbf{l_{\perp}}^2}{y(1-y)}}{8 {\beta
}^2}\right)\nonumber\\
& & \times\left( \frac{-x} {(1-x) \mathbf{q_{\perp}}^2+2
\mathbf{q_{\perp}}\cdot \mathbf{k_{\perp}}} +\frac{-y}
{(1-y)\mathbf{q_{\perp}}^2-2\mathbf{q_{\perp}}\cdot
\mathbf{l_{\perp}}}+ \right.\nonumber\\
& & \left. \frac{x+y-2xy}{2 (1-y) \mathbf{q_{\perp}}\cdot
\mathbf{k_{\perp}}-2(1-x)\mathbf{q_{\perp}}\cdot
\mathbf{l_{\perp}}+ (1-x)(1-y)\mathbf{q_{\perp}}^2} \right) \ ,
\end{eqnarray}
By integrating over the azimuth angles for $\mathbf{k}_{\perp}$
and $\mathbf{l}_{\perp}$ with the integration formula shown in the
Appendix, the above six dimensional integration can be reduced to
four dimensional integration, which can then be dealt with by
numerical calculation with the help of the VEGAS program.

\begin{figure}
\centering
\includegraphics[width=0.50\textwidth]{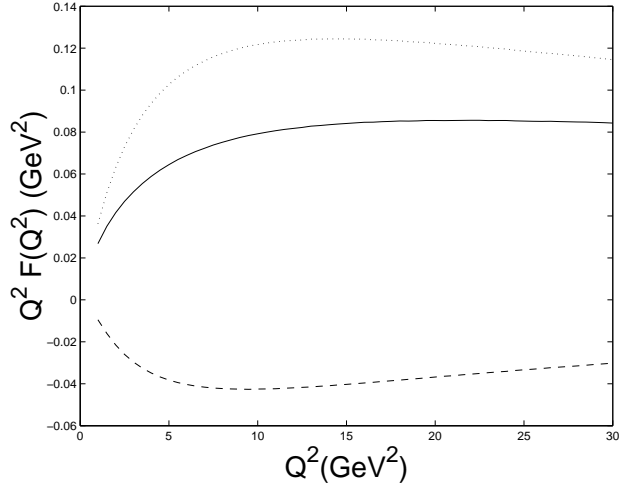}
\caption{The hard contribution to the pion form factor
$Q^2F_{\pi}(Q^2)$. The dotted line stands for the contribution
from the usual helicity ($\lambda_1+\lambda_2=0$) components, the
dashed line stands for the contribution from the higher helicity
($\lambda_1+\lambda_2=\pm 1$) components and the solid line is the
total hard contribution, which is the combined result for all the
helicity components.} \label{hard}
\end{figure}

Integrating over the azimuth angles for $\mathbf{k}_{\perp}$ and
$\mathbf{l}_{\perp}$, we obtain the contribution from the usual
helicity components $(\lambda_1+\lambda_2=0)$,
\begin{eqnarray}\label{hel0f}
F^{(\lambda_1+\lambda_2=0)}_{\pi}(Q^2)&=&\int dx dy d\eta_1
d\eta_2\frac{A^2 \xi_1 {C_F}
\alpha_s(Q^2)|\mathbf{k}_{\perp}||\mathbf{l}_{\perp}|}{32\pi^3 x
y}\exp\left(-\frac{\frac{m^2 +|\mathbf{k_{\perp}}|^2} {x(1-x)} +
\frac{m^2 + |\mathbf{l_{\perp}}|^2}{y(1-y)}}{8 {\beta
}^2}\right)\times\nonumber\\
&
&\!\!\!\!\!\!\!\!\!\!\!\!\!\!\!\!\!\!\!\!\!\!\!\!\!\!\!\!\!\!\!\!\!\!\!\!\!
\left( \frac{x( x + y-1 - 2xy ) } {( 1 - x ) \sqrt{1 - \eta_1^2}}
+\frac{y( x + y -1- 2xy ) } {( 1 - y ) \sqrt{1 - \eta_2^2}} +
\frac{x + y- x^2 - y^2} {( 1- x ) (1-y )\sqrt{1 -\eta_1^2} \sqrt{1
- \eta_2^2}} \right)
\end{eqnarray}
and the contribution from the higher helicity components
$(\lambda_1+\lambda_2=\pm 1)$,
\begin{eqnarray}\label{hel1f}
F^{(\lambda_1+\lambda_2=\pm 1)}_{\pi}(Q^2)&=&-\int dx dy d\eta_1
d\eta_2 \frac{A^2 \xi_3 {C_F}\alpha_s(Q^2)|\mathbf{k}_{\perp}|
|\mathbf{l}_{\perp}|}{64 \pi^3 x y }\exp\left( -\frac{\frac{m^2
+|\mathbf{k_{\perp}}|^2} {x(1-x)} + \frac{m^2 +
|\mathbf{l_{\perp}}|^2}{y(1-y)}}{8 {\beta
}^2}\right)\times\nonumber\\
& & \left(( x + y - 2xy ) \frac{( 1 - \sqrt{1 - \eta_1^2})( 1 -
\sqrt{1 - \eta_2^2})}{(1-x)(1-y)\eta_1\eta_2 \sqrt{1
-\eta_1^2}\sqrt{1 -\eta_2^2}} \right)\ ,
\end{eqnarray}
where $\xi_3=\frac{|\mathbf{k}_{\perp}|
|\mathbf{l}_{\perp}|}{[m^2+\mathbf{k}^2_{\perp}]^{1/2}
[m^2+\mathbf{l}^2_{\perp}]^{1/2}}$,
$|\mathbf{k}_{\perp}|=Q(1-x)\eta_1/2$ and
$|\mathbf{l}_{\perp}|=Q(1-y)\eta_2/2$, with $\eta_{1,2}$ in the
range of $(0,\ 1)$. In Eq.(\ref{hel1f}), there is an overall minus
sign and because the integrand is always positive, we can draw the
conclusion that the higher helicity components will always {\it
suppress} the contribution from the usual helicity components.

With the help of the LC wave function Eq.(\ref{wave}) and its
parameter values shown in Eq.(\ref{dataresult}), we show the pion
form factor with or without the higher helicity components in
Fig.\ref{hard}. One may observe a large {\it suppression} comes
from the higher helicity components for the pion form factor as
compared to the prediction obtained in the original hard
scattering model\cite{mh}. This large suppression was obtained by
Ref.\cite{wk} with a quite different picture. They argued that the
transverse momentum in the quark propagator is of small
contribution (about $15\%$\cite{lis,qpropagator}). The
hard-scattering amplitude, after neglecting the transverse
momentum dependence in the quark propagator, was taken to be
\begin{equation}\label{wang}
T_H^{(\lambda_1+\lambda_2=\pm 1)}=-T_H^{(\lambda_1+\lambda_2=0)}
=-\frac{4g^2C_F}{x_2y_2Q^2+({\bf k}_\perp-{\bf l}_\perp)^2}
\stackrel{Q^2\rightarrow\infty} \approx -\frac{4g^2C_F}{x_2y_2
Q^2}+\frac{4g^2C_F({\bf k}_\perp-{\bf l}_\perp)^2}{(x_2y_2
Q^2)^2}.
\end{equation}
It can be seen from Eq.(\ref{wang}) that the asymptotic
($Q^2\rightarrow \infty$) behaviors of the two helicity states are
directly with {\sl opposite} signs and both states make the
contribution at the order of $1/Q^2$. However it is not a right
argument and it is this factor that causes the asymptotic behavior
of higher helicity contribution is of order $1/Q^4$ other than
$1/Q^2$. In the present work, we have considered the $k_T$
dependence both in the wave function and in the hard scattering
amplitude consistently within the LC PQCD approach, then our
results have a right power behavior for the higher helicity
components' contributions.

\section{a discussion of the soft contribution to the pion form factor}

In the above sections, we have shown that the inclusion of the
higher helicity components {\it suppresses} the hard scattering
contribution at moderate $Q^2$. In order to compare our
predictions with the present experimental data, we need to know
the contribution from the soft part. Since this part is model
dependent and is still under progress\cite{jkw,kcj,sch,amn}, as an
example, we consider the soft contribution to the pion form factor
with the model LC wave function shown in Eq.(\ref{wave}) and study
the different helicity components' soft contribution to the pion
form factor separately.

For the soft part contribution, we have\cite{dy}
\begin{equation}\label{gensoft}
F^s_{\pi^+}(Q^2)=\int^1_0 dx \int \frac{d^{2}{\bf
k}_{\perp}}{16\pi^3}\sum_{\lambda,\lambda^{\prime}}\Psi^*(x,{\bf
k_{\bot}},\lambda)\Psi(x,{\bf k'_{\bot}},\lambda^{\prime})+\cdots,
\end{equation}
where $\lambda$, $\lambda^{\prime}$ are the helicities of the wave
function respectively, and the first term is the lowest order
contribution from the minimal Fock-state and the ellipses
represent those from higher Fock states, which are down by powers
of $1/Q^2$ and by powers of $\alpha_s$.

Taking the LC wave function as is shown in Eq.(\ref{wave}), we
obtain
\begin{equation}
\label{softpion} F^s_{\pi^+}(Q^2)=\int^1_0 dx \int \frac{d^{2}{\bf
  k}_{\perp}}{16\pi^3}\frac{m^2+\mathbf{k}_{\perp}
  \cdot\mathbf{k'}_{\perp}}{\sqrt{m^2+\mathbf{k}_{\perp}^2}
  \sqrt{m^2+\mathbf{k'}_{\perp}^2}}\times A^2
  \exp\left(-\frac{{\bf k}_{\perp}^2+m^2}{8{\beta}^2x(1-x)}-
  \frac{{\bf k'}_{\perp}^2+m^2}{8{\beta}^2x(1-x)}\right),
\end{equation}
where ${\bf k'}_{\perp}={\bf k}_{\perp}+(1-x){\bf q}_{\perp}$ for
the final state LC wave function when taking the Drell-Yan-West
assignment\cite{dy}. We proceed to integrate the transverse
momentum $\mathbf{k}_{\bot}$ in Eq.(\ref{softpion}) with the help
of the Schwinger $\alpha-$representation method,
\begin{equation}\label{schwinger}
\frac{1}{A^{\kappa}}=\frac{1}{\Gamma(\kappa)}\int_0^{\infty}
\alpha^{\kappa-1}e^{-\alpha A}d\alpha\ .
\end{equation}
Doing the integration over $\mathbf{k}_{\perp}$, we obtain
\begin{eqnarray}
F^s_{\pi^+}(Q^2)&=&\int^1_0 dx\int^\infty_0 d\lambda
\frac{A^2}{128\pi^2 ( 1 + \lambda)^3}\exp\left(-\frac{8m^2{( 1 +
\lambda) }^2 + Q^2(1-x)^2( 2 + \lambda( 4 + \lambda))}{32 (1-
x ) x\beta^2 ( 1 + \lambda) }\right)\nonumber\\
& & \times\left( I_0\left(\frac{Q^2( x-1) \lambda^2} {32x\beta ^2(
1 + \lambda)}\right)\bigg( 32( 1 - x ) x\beta^2( 1 + \lambda)
 -Q^2( 1 - x )^2 ( 2 + \lambda ( 4 +\lambda )) \right.\nonumber\\
& & \left. + 8m^2(1 +\lambda)^2\bigg) - I_1\left(\frac{Q^2(x -1)
{\lambda }^2} {32x\beta^2( 1 + \lambda) }\right)Q^2(1-x)^2
\lambda^2 \right),
\end{eqnarray}
where the $I_n\,\,(n=0,1)$ is the modified Bessel function of the
first kind. After taking the expansion in the small $Q^2$ limit,
we obtain the probability,
\begin{eqnarray}
P_{q\bar{q}}&=& F^s_{\pi^+}(Q^2)|_{Q^2=0}=\int
dx\frac{d^2\mathbf{k}_{\perp}}
{(16\pi)^3}|\Psi(x,\mathbf{k}_{\perp})|^2 \nonumber\\
&=&\int^1_0 dx\int^\infty_0 d\lambda \frac{A^2} {16\pi^2 ( 1 +
\lambda)^2}\exp\left(\frac{m^2( 1 + \lambda)}{4 (x-1) x\beta^2
}\right)\times\Bigg(m^2(1 +\lambda)+4x(1-x)\beta^2\Bigg)
\end{eqnarray}
and the charged mean square radius $\langle
r^2_{\pi^+}\rangle^{q\bar{q}}$\cite{cardar2},
\begin{eqnarray}\label{radius}
\langle r_{\pi^+}^2 \rangle^{q\bar{q}}
&\approx&-6\left.\frac{\partial
F^s_{\pi^+}(Q^2)}{\partial Q^2}\right|_{Q^2=0}\nonumber\\
&=&\int^1_0 dx\int^\infty_0 d\lambda\frac{3A^2}{256 {\pi
}^2x{\beta }^2{( 1 + \lambda) }^3}\exp\left(-\frac{m^2( 1 +
\lambda) } {4(1-x) x{\beta }^2}\right)( 1-x )( 2 + 4\lambda+
{\lambda }^2 )\nonumber\\
& &\times\Bigg( 8( 1-x) x{\beta }^2 +m^2( 1 + \lambda)\Bigg) \ .
\end{eqnarray}
In the above two equations, one may observe that the terms in the
big parenthesis that are proportional to $m^2$ come from the
ordinal helicity components, while the remaining terms in the big
parenthesis are from the higher helicity components.

\begin{figure}
\centering
\begin{minipage}[c]{0.45\textwidth}
\centering
\includegraphics[width=2.9in]{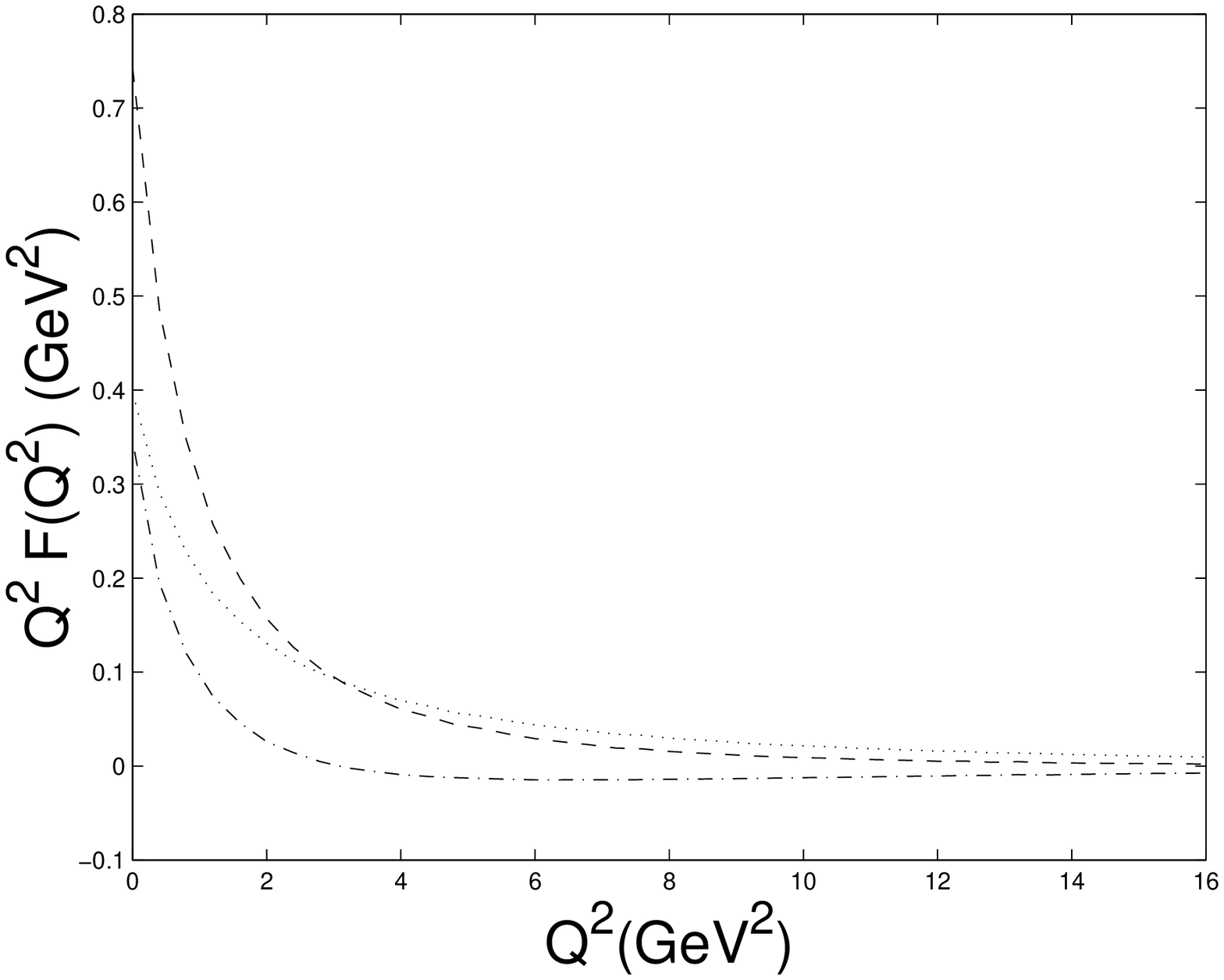}
(a)
\end{minipage}%
\hspace{0.2in}
\begin{minipage}[c]{0.45\textwidth}
\centering
\includegraphics[width=2.9in]{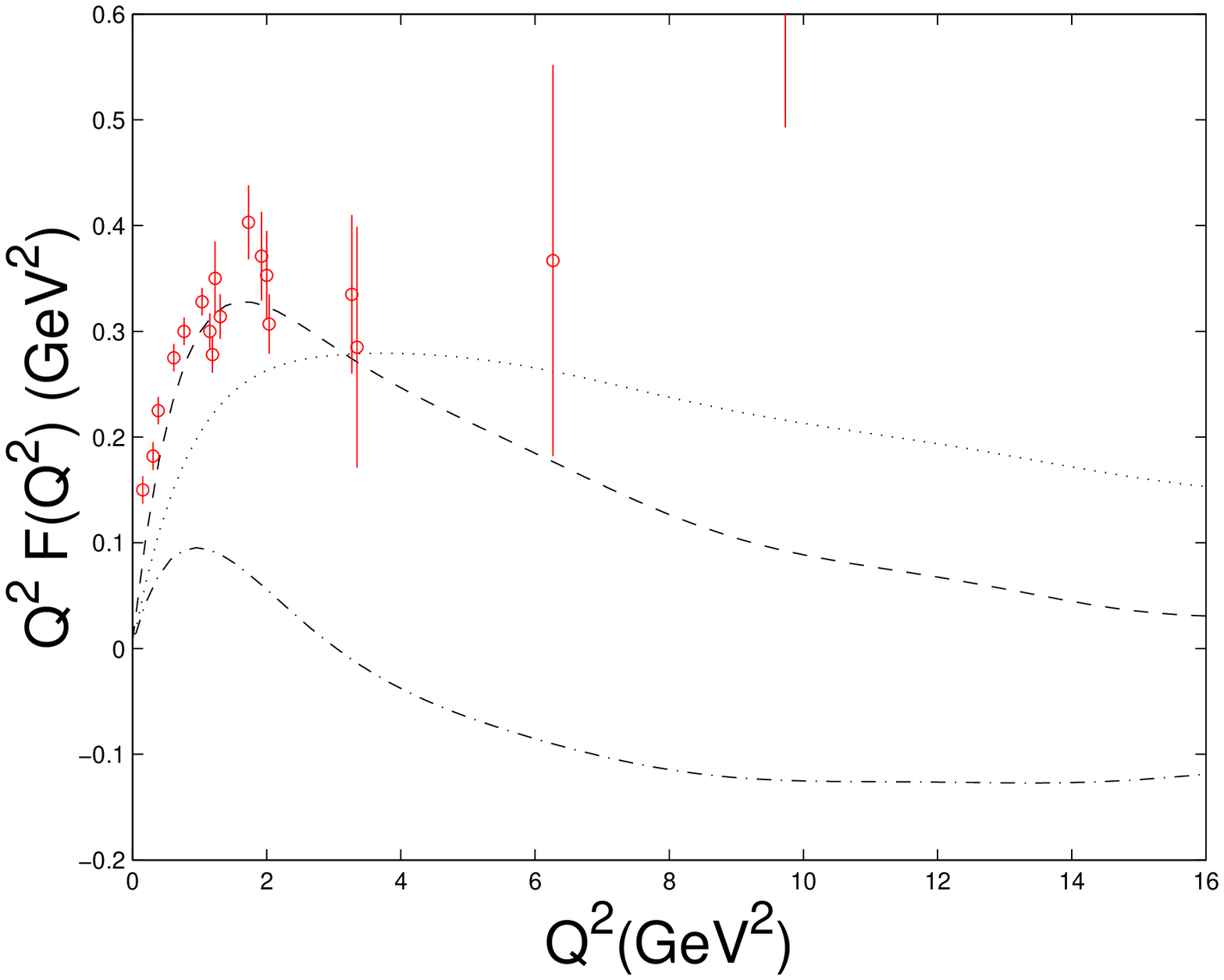}
(b)
\end{minipage}
\caption{The soft contribution to the pion form factor. Left is
for the pion form factor $F_{\pi}(Q^2)$, while the right is for
$Q^2F_{\pi}(Q^2)$, where the contribution comes from all the
helicity components are shown in dashed line, the contribution
from the ordinal helicity component $\lambda_1 +\lambda_2=0$ is
shown in dotted line and the contribution from the higher helicity
components $\lambda_1 +\lambda_2=\pm 1$ is shown in dash-dot line.
The experimental data is taken from\cite{cjb}.} \label{form}
\end{figure}

The parameters in the wave function can be determined by several
reasonable constraints~\cite{hms}. Two constraints can be derived
from $\pi\to\mu\nu$ and $\pi^0\rightarrow \gamma\gamma$ decay
amplitude~\cite{bhl}:
\begin{equation}
\label{Aconstrain} \int^1_0 dx \int \frac{d^{2}{\bf
k}_{\perp}}{16\pi^3}\Psi(x,{\bf k}_{\perp}) =f_{\pi}/(2\sqrt{3}),
\end{equation}
and
\begin{equation}
\label{Bconstrain} \int^1_0 dx \Psi(x,{\bf k}_{\perp}=0)
=\sqrt{3}/f_{\pi},
\end{equation}
where $f_{\pi}$ is the pion decay constant: $ \langle
0|\bar{q}(0)\gamma^+\gamma_5 q(0)|P\rangle=if_{\pi}P^+$, the
experimental value of which is $92.4\pm 0.25 MeV$\cite{pdg}.
Experimentally the average quark transverse momentum of pion
$\langle\mathbf{k}_{\perp}^2\rangle_{\pi}$ is approximately of the
order $(300 MeV)^2$\cite{metcalf}. The quark transverse momentum
of the valence state in the pion is defined as
\begin{equation}\label{Cconstrain}
\langle \mathbf{k}_{\perp}^2\rangle_{q\bar{q}}=\int
dx\frac{d^2\mathbf{k}_{\perp}} {(16\pi)^3}|
\mathbf{k}_{\perp}^2|\frac{|\Psi(x,\mathbf{k}_{\perp})|^2}
{P_{q\bar{q}}},
\end{equation}
and it should be larger than $\langle\mathbf{k}_{\perp}^2
\rangle_{\pi}$. We thus could require that
$\sqrt{\langle\mathbf{k}_{\perp}^2\rangle_{q\bar{q}}}$ has the
value of a few hundreds MeV, serving as the third constraint.
Using the constraints and the model wave function Eq.(\ref{wave}),
we obtain,
\begin{equation}\label{dataresult}
m=310MeV\,\,;\,\,\beta=396MeV;\,\, A=0.050MeV^{-1} \ ,
\end{equation}
for $\langle \mathbf{k}^2_{\perp}\rangle \approx (367 MeV)^2$. And
by using the above parameters, we obtain
\begin{eqnarray}
\langle r^2_{\pi^+}\rangle^{q\bar{q}} &=& 0.216 fm^2,\\
\label{wholenor} P_{q\bar{q}}&=&
P^{(\lambda_1+\lambda_2=0)}_{q\bar{q}}
+P^{(\lambda_1+\lambda_2=\pm 1)}_{q\bar{q}}=0.744\ .
\end{eqnarray}
The value of $\langle r^2_{\pi^+}\rangle^{q\bar{q}}$ is in nice
agreement with the ones obtained in Ref.\cite{huang,cardarelli}.
In fact, we have used the same monopole ansatz see in
Ref.\cite{cardarelli,cardar2}. It is shown that the valence quark
radius is smaller than the experimental value of the pion charged
radius ($(0.671\pm 0.008 fm)^2$\cite{pdg}). Therefore the valence
portion of a hadron is more compact than the hadron radius. For
the probability of finding the valance states in the pion, we have
$(P_{q\bar{q}}^{(\lambda_1+\lambda_2=0)}=0.398)$ for the usual
helicity components and $(P_{q\bar{q}}^{(\lambda_1+\lambda_2=\pm
1)}=0.346)$ for the higher helicity states, which show that the
higher helicity components have the same importance as that of the
usual helicity components. It has been shown that even though we
have added the contributions from the higher helicity states, the
probability of finding the minimal $q\bar{q}$ Fock state in pion
is still less than unity, i.e. $(P_{q\bar{q}}=0.744)<1$. This is
shown clearly in Fig.\ref{form}.(a), so it is necessary to take
the higher Fock states and the higher twist terms into
consideration to give a full understanding of the pion form factor
at the energy region $Q^2\rightarrow 0$. It should be noticed that
if one normalizes the valence Fock state to unity without
including the higher helicity components, then the soft and hard
contributions from the valence state can be enhanced and become
important inadequately.

The result for the soft contribution to the pion form factor is
shown in Fig.\ref{form}. From Fig.\ref{form}.(b), one may observe
a quite different behavior from that of the hard contribution for
the higher helicity components ($\lambda_1+\lambda_2=\pm 1$). In
the energy region $Q^2\lesssim 1GeV^2$, the higher helicity
components give a large enhancement (the same order contribution)
to usual helicity ($\lambda_1+\lambda_2=0$) components and after
that the higher helicity components' contributions will decrease
with the increasing $Q^2$. At about $Q^2\sim 4GeV^2$, the higher
helicity components' contributions become negative and as a
result, the net soft contribution will then decrease fast with the
increasing $Q^2$, which tends to zero at about $Q^2\sim 16GeV^2$.

\begin{figure}
\centering
\includegraphics[width=0.50\textwidth]{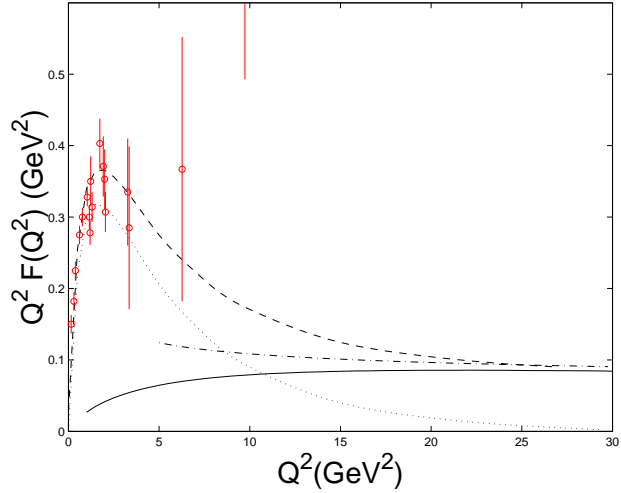}
\caption{The combined results for the pion form factors
$Q^2F_{\pi}(Q^2)$. The solid line stands for the contribution from
the hard part, the dotted line stands for the contribution from
the soft part, the dashed line is the total Pion form factors and
the dash-dot line is the usual asymptotic result. The experimental
data are taken from \cite{cjb}.} \label{total}
\end{figure}

We show the combined results that come from the hard scattering
part and from the soft part for the pion form factors
$Q^2F_{\pi}(Q^2)$ in Fig.\ref{total}, where for comparison, the
experimental data \cite{cjb} and the well-known asymptotic
behavior for the leading twist pion form factor have also been
shown. It it shown that the soft contribution is less important as
$Q^2> a\;few\;GeV^2$, since we have taken the correct
normalization condition Eq.(\ref{wholenor}) and considered the
suppression effect from the higher helicity components. One may
observe that our present result for the pion form factor is lower
than the experimental data, it is reasonable since we have not
taken the higher twist effects and the higher order corrections
into consideration. The next-to-leading order correction will give
about $\sim 20-30\%$\cite{field} extra contribution to the pion
form factor, while the twist-3 contribution is comparable with the
leading twist contribution in a large intermediate energy region
($\sim 1-40GeV$) \cite{twist}\footnote{The twist-3 contribution is
model dependent, if we take the wave function with a better end
point behavior, then the twist-3 contribution can be greatly
suppressed\cite{huangwu}.}.

\section{Summary and conclusion}

In this paper, the transverse momentum effects and the higher
helicity components' contributions to the pion form factor are
systematically studied based on the LC framework and the $k_T$
factorization formalism. Both collinear and $k_T$ factorization
are the fundamental tools for applying PQCD to the pion form
factor since they can separate the calculable perturbative
contributions from the non-perturbative parts that can be absorbed
into the bound-state wave functions. The $k_T$ factorization
theorem has been widely applied to various processes and the $k_T$
factorization theorem for exclusive processes in PQCD has been
proved by M. Nagashima and H.N. Li. Thus it provides a scheme to
take the dependence of the parton transverse momentum $k_T$ into
account. Ref.\cite{lis} shows that the end-point singularity can
be cured by resuming the resultant double logarithms
$\alpha_s\ln^2 k_T$ into a Sudakov form factor and then the PQCD
analysis can make sense. In fact, the Sudakov effects have a small
effects for the pion form factor in the region where experimental
results are available. We note that there are $k_T$ dependence in
the wave function in the $k_T$ factorization and it generates much
larger effects than the Sudakov suppression to the hard scattering
amplitude in the present experimental $Q^2$ region. Our results
show that it is substantial to take $k_T$ dependence in the wave
function into account.

The light cone formalism provides a convenient framework for the
relativistic description of the hadron in terms of quark and gluon
degrees of freedom, and the application of PQCD to exclusive
processes has mainly been developed in this formalism. In the
present paper, we have given a consistent treatment of the pion
form factor within the LC PQCD framework, i.e. both the wave
function and the hard interaction kernel are treated within the
framework of LC PQCD. Taking into account the spin space Wigner
rotation, one may find that there are higher-helicity components
($\lambda_1+\lambda_2=\pm 1$) in the LC spin-space wave function
besides the usual-helicity components ($\lambda_1+\lambda_2=0$).
We have studied the higher helicity components' contributions to
the hard part and the soft part of the pion form factor by using
the light cone PQCD approach with the parton's transverse momentum
$k_T$ included. We find that the asymptotic behavior of the
hard-scattering amplitude for the higher-helicity components
including the transverse momentum in the quark propagator is of
order $1/Q^4$ which is the next to leading order contribution
compared with the contribution coming from the ordinary helicity
component, but it can give sizable contribution to the pion form
factor at the intermediate energies.

In order to compare our predictions with the experimental data, we
need to know the contribution from the soft part. As an example,
we have considered the soft contribution to the pion form factor
with a reasonable wave function in the LC framework. Our results
show that the soft contribution from the higher helicity
components has a quite different behavior from that of the hard
scattering part and has the same order contribution as that of the
usual helicity ($\lambda_1+\lambda_2=0$) components in the energy
region ($Q^2\lesssim 1GeV^2$). As $Q^2>1GeV^2$, the higher
helicity components' contributions will decrease with the
increasing $Q^2$. At about $Q^2\sim 4GeV^2$, the higher helicity
components' contributions become negative and as a result the net
soft contribution to the pion form factor will then decrease with
the increasing $Q^2$, which tends to zero at about $Q^2\sim
16GeV^2$. Thus the soft contribution is less important in the
intermediate energy region. Although the soft contribution is
purely non-perturbative and model-dependent, our results show that
the calculated prediction for the pion form factor should take the
$k_T$ dependence in the soft and hard parts into account beside
including the higher order contribution. Therefore one needs to
keep the transverse momentum in the next leading order corrections
and to construct a realistic $k_T$ dependence in the hadronic wave
function in order to derive more exact prediction to the pion form
factor in $k_T$ factorization.

\begin{center}
\section*{Acknowledgements}
\end{center}

We would like to thank Drs F.G. Cao, J. Cao, H.N. Li and B.Q. Ma
for useful discussions. This work was supported in part by the
Natural Science Foundation of China (NSFC).\\

\appendix
\section{integration formula}
\label{app:integration}

The error function is defined as
\begin{equation}
Erf(x)=\frac{2}{\sqrt{\pi}}\int_0^{x} e^{-t^2} dt \ .
\end{equation}
An important property for the error function is
$\lim_{x\rightarrow \infty} Erf(x)=1$.

Second we list some useful formula that are needed for integrating
over the azimuth angle of the momenta $\mathbf{k}_{\perp}$ and
$\mathbf{l}_{\perp}$ and then reduce the integration dimension
from six to four. And the remaining four dimensional integration
can be done numerically.

By using polarization coordinate, we have
\begin{equation}
[d\mathbf{k}_{\perp}^2][d\mathbf{l}_{\perp}^2]= kl dk dl d\theta
d\rho/(16\pi^3)^2,
\end{equation}
where $k$, $l$ and $\theta$, $\rho$ are the module  and azimuth
angle of $\mathbf{k}_{\perp}$ and $\mathbf{l}_{\perp}$
respectively. By using the following formula, the integration over
the azimuth angle can be done analytically.
\begin{eqnarray}
f_{1}(A,B)&=&\int_0^{2\pi}\frac{d\theta}{A+ B \cos(\theta)}
=\frac{2\pi}{\sqrt{(A+B)(A-B)}}\ ,\\
f_{2}(A,B)&=&\int_0^{2\pi}\frac{\cos(\theta)d\theta}{A+B\cos(\theta)}
=\frac{2\pi}{B}\left(1-\frac{A}{\sqrt{(A+B)(A-B)}}\right)\ ,\\
f_{3}(A,B)&=&\int_0^{2\pi}\frac{\sin(\theta)d\theta}{A+ B
\cos(\theta)} =0 \, \\
f_{4}(A,B)&=&\int_0^{2\pi}\int_0^{2\pi}\frac{\cos(\theta-\rho)
d\theta
d\rho}{A+B\cos(\theta)}=0\, \\
f_{5}(A,B,C)&=&\int_0^{2\pi}\int_0^{2\pi}\frac{\cos(\theta-\rho)
d\theta d\rho}{A+B\cos(\theta)+C\cos(\rho)}\ ,
\end{eqnarray}
where $A$, $B$ and $C$ are functions that are free from $\theta$
$\rho$. The result for the function $f_5$ is very complicated and
for simplicity it's explicitly form will not be listed here.
However, by adding a small component ($BC\cos(\theta)\cos(\rho)$)
(for the integration we need to deal with, we have $BC<<A$, which
corresponding to $\mathbf{k}_{\perp}\cdot\mathbf{l}_{\perp}<<
\mathbf{q}^2_{\perp}$), it can be solved approximately,
\begin{equation}
f_{5}(A,B,C)\approx
\int_0^{2\pi}\int_0^{2\pi}\frac{\cos(\theta-\rho) d\theta
d\rho}{(A+B\cos(\theta))(A+C\cos(\rho))} =f_{2}(A,B)f_{2}(A,C)\ .
\end{equation}
There one may notice that under the present approximation, the
actual azimuth angle ,i.e. $\alpha$, for $\mathbf{q}_{\perp}$ will
not affect the final integrated results due to the fact that after
integration over $\theta$ and $\rho$, it will always accompanied
by a factor $(\cos(\alpha)^2+\sin(\alpha)^2)\equiv 1$.

After integrating over the azimuth angle, we can change the
integration over the radius of $\mathbf{k}_{\perp}$ and
$\mathbf{l}_{\perp}$ to two dimensionless variables $\eta_1$ and
$\eta_2$ that are within the range of $(0,1)$ through the relation
\begin{equation}
|\mathbf{k}_{\perp}|=Q(1-x)\eta_1/2\ , \ \ \ \
|\mathbf{l}_{\perp}|=Q(1-y)\eta_2/2.
\end{equation}
The relation is so choosing as to insure that all the quantities
in the radical sign obtained by doing the azimuth angle
integration are always positive.

\end{document}